# Filtered Fictitious Play for Perturbed Observation Potential Games and Decentralised POMDPs

**Archie C. Chapman**
University of Sydney Business School
University of Sydney
Sydney 2006, Australia

**Simon A. Williamson**
School of Information Systems
Singapore Management University
Singapore 178902

**Nicholas R. Jennings**
Electronics and Computer Science
University of Southampton
Highfield, SO17 1BJ, UK

## Abstract

Potential games and decentralised partially observable MDPs (Dec–POMDPs) are two commonly used models of multi–agent interaction, for static optimisation and sequential decision–making settings, respectively. In this paper we introduce *filtered fictitious play* for solving repeated potential games in which each player's observations of others' actions are perturbed by random noise, and use this algorithm to construct an online learning method for solving Dec–POMDPs. Specifically, we prove that noise in observations prevents standard fictitious play from converging to Nash equilibrium in potential games, which also makes fictitious play impractical for solving Dec–POMDPs. To combat this, we derive filtered fictitious play, and provide conditions under which it converges to a Nash equilibrium in potential games with noisy observations. We then use filtered fictitious play to construct a solver for Dec–POMDPs, and demonstrate our new algorithm's performance in a box pushing problem. Our results show that we consistently outperform the state–of–the–art Dec–POMDP solver by an average of 100% across the range of noise in the observation function.

## 1 Introduction

Increasingly, complex real-world problems are being tackled by teams of autonomous agents. Now, in artificial intelligence, the problem of controlling these agents is often framed as a problem of distributed optimisation (e.g. [11]), and one general method of optimisation is for each agent to iteratively choose a better reply to the actions of the others, given its beliefs. This paper focuses on one particular iterative algorithm, *fictitious play* [5], and its application to controlling teams of agents in single state and sequential decision–making problems. However, a common feature of the domains that agent systems are deployed in is uncertainty about the actions of other team members, and this can prevent algorithms like fictitious play from converging.

The specific type of uncertainty we consider in this paper is when agents cannot accurately infer (either via sensors or communication) the actions that were just taken by the others in the team. We call this scenario *perturbed observations* to reflect the fact that it is the players' observations, not their actions, that are noisy. As such, it requires the players to incorporate the possibility of incorrect observations into their learning procedures. We envisage this model applying to situations in which the players have on–board payoff evaluation, which means that the joint action that was just played cannot be inferred from the payoff an agent receives. This feature arises in many scenarios, such as: (i) mobile multi–robotics where agents share position information using noisy odemetry — coordination is difficult when robots cannot share accurate positioning information, (ii) unmanned aerial vehicle (UAV) target tracking, in which individual UAVs need to estimate what the others are currently tracking — images of a target may be needed from many positions but a UAV can only approximately infer the viewing angle of others, and (iii) distributed sensor optimisation problems, where there are often significant costs on communication (e.g. due to power restrictions), so sensors use noisy communication channels to coordinate.

Against this background, *potential games* [12] are an important class of games that can be used as a design template for agent–based distributed optimisation problems. This class contains as special cases several important models of multi–agent interaction, including: team or common–interest games [18], where players have the same utility function, and; marginal contribution games [11], where the players are each rewarded with their marginal contribution to a global target function. In more detail, a potential game can be constructed from a global target function by distributing the system's control variables among a set of players (or agents), and *aligning* each player's utility function with the system–wide goals. That is, payoffs are derived so that a player's payoff for changing strategy increases if and only if the global reward also increases (as in [20]). If the players' utilities are aligned with the global target function, then the global target function is a *potential function* for the game, and the game is a potential game. (The models above are two examples of games in which players' utilities are aligned with the global target function.) Importantly, if the agents' utilities and the global reward are aligned, then the pure Nash equilibria (NE) of the resulting game correspond to the local optima of the global target function, and, furthermore, these can be computed in a distributed fashion using an iterative algorithm such as fictitious play. However, perturbations to the agents' observations, which may cause an agent to incorrectly record the actions of others, may disrupt the convergence of such iterative algorithms.

Within this context, the first contribution of this paper is to identify precisely the conditions under which noise in action observations prevents fictitious play from converging in potential games.[1] In more detail, we show that if action observations are noisy, some (or all) strict NE are removed from the set of possible stationary points of any standard fictitious play process; in some cases, the global optimum for a team problem may be removed from the set of stationary points of fictitious play, which is clearly problematic. Our second contribution, then, is to derive a generalised version, called *filtered fictitious play* (FFP), which addresses the above shortcoming. We derive a method for integrating a noise filtering process with the belief revision of fictitious play, and derive a specific version of FFP based on a Bayesian filter. In potential games with perturbed action observations, we prove that FFP converges to a pure NE in many cases where standard fictitious play fails due to the noise (but would have converged in its absence).

Building on these results, we then move on to consider *partially observable stochastic games* [7]. Stochastic games comprise several states or stage games, with agents' actions causing transitions between these states. Partially observable stochastic games (POSGs) have the additional complication that the agents may be uncertain of what the current state actually is, and as such, POSGs capture multi–agent sequential decision–making under uncertainty. With this in mind, a stochastic game can also possess a potential function.[2] One prominent example of this is called a *team Markov game* [10], which has as a potential function the team objective function that is to be optimised by the agents. As such, fictitious play could be used to solve these games, and indeed, a version of a related algorithm called adaptive play was used by [18] to solve team Markov games. Fictitious play presents a useful direction for solving POSGs, since other agents can be grouped together and effectively regarded as a single entity which the deliberating agent must respond to. This makes it naturally distributed and of lower computational complexity compared to distributed planning approaches. However, it is not immediately clear how to extend fictitious play from stochastic games to the more general class of POSGs, or even the simplest subclass of POSGs that directly generalise team Markov games, namely *decentralised partially observable MDPs* (Dec–POMDPs). This is because standard fictitious play is not equipped to deal with uncertain action observations. Specifically, action observation perturbations prevent fictitious play from correctly identifying the state, which induces errors in policy valuations that are propagated through the candidate solutions. The third contribution of this paper, then, is to show how FFP can be extended to solving POSGs in which uncertainty in the actions of the other agents makes it hard for the agent to track which stage game (state) it is currently in. We use FFP as a component in a new online reinforcement learning algorithm based on local search. Experimental results in a classic cooperative box pushing Dec–POMDP with noisy action observations show that it significantly outperforms the state of the art when allowed to learn for the same length of time as the state-of-the-art algorithm's runtime.

**Related work:** Several researchers have investigated reinforcement learning techniques for computing NE in stochastic games, and general–purpose algorithms such as R–max, OAL and WoLF have been derived (for a review and empirical comparison, see [2]). However, all these algorithms rely on perfect observations of the actions of the other agents. On the other hand, algorithms have also been developed for POSGs (e.g. [1], or the current state of the art, *Point Based Policy Generation*, PBPG [21]), but these approaches are approximate and not proven to converge to a NE. This is because, in general, solving POSGs offline for a decentralised policy is a NEXP–complete problem [4]. Now, there are several possible ways to combat this problem, such as finding a centralised policy offline (PSPACE) and executing it online using a communication protocol [14]. Unfortunately, deriving the communication protocol generally is again a hard problem. In response, approaches such as [16] use heuristic communication protocols during execution, however they have no performance guarantees. Now, agents can also attempt to optimise a POSG online using local techniques (e.g. [19]), which are scalable as the entire problem is not solved, but these approaches also are not guaranteed to find a NE. Given this, we aim to develop a scalable method for learning NE in partially observable domains. Finally, we could have chosen to ignore the other agents entirely by treating them as a noisy part of the environment, and then apply standard reinforcement learning techniques, as in [8]. In practice, however, such approaches learn very slowly, because the rewards obtained are extremely noisy — and possibly non–stationary, due to changes in the underlying choice of actions induced by noisy observations. In contrast, using fictitious play to respond to the other agents' actions removes the possibility of such non–stationarity in rewards arising.

---

[1] Fictitious play also solves other classes of games, such as two–player zero–sum games, but in this paper we focus on potential games because of their connection to optimisation problems.

[2] Potential games are typically characterised in terms of single state games, but the definition is extended to stochastic games by defining a potential over the joint policy space. If such a potential can be constructed, then the stochastic game is a potential game.

**Paper structure:** Section 2 introduces potential games, and reviews the analytical tools used: *p*–dominance and results for generalised weakened fictitious play (GWFP) (a broad class of fictitious play processes). In Section 3 we introduce games with perturbed observations, and explain why GWFPs do not necessarily converge in these games. This motivates Section 4, where we show how to integrate a noise filtering process within the belief revision of GWFP, derive a specific version of FFP based on a Bayesian filter, and prove that FFP converges to pure NE in many games with noisy action observations where GWFP does not. We also numerically compare FFP with a standard GWFP to demonstrate its benefits. In Section 5, we derive an algorithm for solving POSGs, and demonstrate its performance in a standard Dec–POMDP. Section 6 concludes.

## 2 Preliminaries

This section first describes noncooperative games, potential games and the concept of *p*–dominant NE. We then introduce GWFP processes and show how they are analysed using stochastic approximations and differential inclusions.

### 2.1 Noncooperative Games and Potential Games

We consider repeated play of a finite noncooperative game, $\Gamma = \langle N, \{A_i, r_i\}_{i \in N} \rangle$, consisting of a set of players $N = \{1, \ldots, n\}$, and for each $i \in N$, a finite set of (pure) *actions* $A_i$, with joint action space $A = \times_{i=1}^{N} A_i$, and a *reward function* $r_i : A \to \mathbb{R}$. A player's reward function specifies its ranking over all joint action profiles, $a \in A$, also called *outcomes*. Players can choose to play an action according to a lottery, known as a *mixed strategy*. This is a probability distribution over pure actions $A_i$, such that $\pi_i \in \Delta_i$, the set of distributions over $A_i$. The rewards of the mixed extension of the game are given by the expected value of $r_i$ under all players' joint independent lottery $\pi \in \times_{i \in N} \Delta_i$ over $A$:

$$r_i(\pi) = \sum_{a \in A} \left( \prod_{j \in N} \pi_j(a_j) \right) r_i(a). \quad (1)$$

We use the notation $a = (a_i, a_{-i})$ where $a_{-i}$ is the joint action chosen by all players other than $i$, and $\pi = (\pi_i, \pi_{-i})$ where $\pi_{-i}$ is the joint independent lottery similarly.

An agent's goal is to maximise its reward, and a *best response correspondence*, $b_i(\pi_{-i})$, is the set of $i$'s best strategies, given the strategies of the other players:

$$b_i(\pi_{-i}) = \{\pi_i \in \Delta_i : r_i(\pi_i, \pi_{-i}) = \max_{\pi'_i \in \Delta_i} r_i(\pi'_i, \pi_{-i})\}$$

Stable points are characterised by the set of *Nash equilibria*, which are defined as those joint strategy profiles, $\pi^*$:

$$r_i(\pi_i^*, \pi_{-i}^*) - r_i(\pi_i, \pi_{-i}^*) \geq 0 \quad \forall \pi_i, \forall i. \quad (2)$$

That is, in a NE, $\pi_i^* \in b_i(\pi_{-i}^*)$. Building on the NE condition, a *strict* NE is one in which the player is not indifferent between the equilibrium action and any other action (i.e. the inequality in (2) is instead strict), and so must be a pure strategy equilibrium..

We can also weaken a best response, to define a δ–best response and the associated δ–NE, which are useful concepts in our analysis. First, a δ-*best response correspondence*, $b_i^\delta(\pi_{-i})$ is the set of strategies that come within δ of maximising $i$'s reward, conditional on other players' strategies:

$$b_i^\delta(\pi_{-i}) = \{\pi_i \in \Delta_i : r_i(\pi_i, \pi_{-i}) \geq \max_{\pi'_i \in \Delta_i} r_i(\pi'_i, \pi_{-i}) - \delta\}.$$

Then, a strategy profile $\pi^*$ is an δ–NE if $\pi_i^* \in b_i^\delta(\pi_{-i}^*) \ \forall i$.

We now define a specific form of a δ–best response, in which an agent plays (or its opponents observe) a best response with probability $1 - \varepsilon$ and a random non–best response with probability ε. We call such a mixed strategy an ε–best response, $\tilde{b}_i^\varepsilon(\pi_{-i})$, and it places the following probabilities of selection (or observation) on each action $a_i \in A_i$:

$$\tilde{b}_{i,a_i}^\varepsilon(\pi_{-i}) = \begin{cases} \frac{1-\varepsilon}{|b_i(\pi_{-i})|} & \text{if } a_i \in b_i(\pi_{-i}), \\ \frac{\varepsilon}{|A_i| - |b_i(\pi_{-i})|} & \text{otherwise,} \end{cases}$$

Note that an ε–best response returns an expected payoff that is within $\delta = \varepsilon(max_{a \in A}[r_i(a)] - min_{a \in A}[r_i(a)])$ of a best response, and furthermore $\varepsilon \to 0$ implies $\delta \to 0$. Also, note that in the remainder of the paper, when we refer to a specific value, e.g. a 0.8–best response, we will always be referring to an ε–best response (δ–best responses are used for analysis only).

The class of *potential games* is characterised as those games that admit a function specifying the participants' joint preference over outcomes [12], known as a potential function. This is a function on the joint action space $A$ such that the difference in the potential induced by a unilateral deviation of action equals the change in the deviator's reward, i.e. $\forall i \in N$, $\forall a_i, a'_i \in A_i$, and $\forall a_{-i} \in A_{-i}$:

$$P(a_i, a_{-i}) - P(a'_i, a_{-i}) = r_i(a_i, a_{-i}) - r_i(a'_i, a_{-i})$$

Importantly, the local optima of the potential function are NE of the game; that is, the potential is locally maximised by self-interested agents in a system. In order to highlight the connections to distributed optimisation, observe that if the players' rewards are aligned with the global target function (i.e. an increase in a player's reward improves the system reward), then the global target function is a potential function for the game. This, in turn, implies that the (pure) NE of the game correspond to the local optima of the target function. This is a very useful property, because it implies that the local optima of the target function are stable.

### 2.2 p–Dominant Equilibria

We make use of the concept of *p*–dominance [13] to explain the effect of perturbed action observations on GWFP

learning processes. The *p*–dominance criterion is a stability concept. Specifically, a strict pure strategy NE is *p*–dominant if each agent's action is a best response to any joint mixed strategy placing at least probability *p* on the other agents playing their pure strategies in the equilibrium.

Formally, in a *p*–dominant NE, each agent's action must be a best response to any joint mixed profile placing least *p* on the pure equilibrium profile. Define:

$$\bar{b}^p_{-i,a^*} = \{\pi_{-i} \in \Delta_{-i} : (\pi_{-i}(a^*_{-i}) \geq p)\}$$

as the set of joint profiles of *i*'s opponents that place at least probability *p* on a particular equilibrium profile $a^*$. Then, a pure NE $a^*$ is called *p*–dominant if the following holds for all $\bar{\pi}^p_{-i,a^*} \in \bar{b}^p_{-i,a^*}$:

$$r_i(a^*_i, \bar{\pi}^p_{-i,a^*}) - r_i(\pi_i, \bar{\pi}^p_{-i,a^*}) \geq 0 \quad \forall \pi_i \in \Delta_i, \forall i \in N. \quad (3)$$

An intuitively appealing interpretation of *p*–dominance is that it defines a spectrum of stability between strict NE at $p = 1$ to dominant strategy equilibria at $p = 0$.

Alongside *p*–dominant equilibria, we will refer to the *minimum p–dominant* equilibrium in a game, which is the NE with the lowest value of *p* for which (3) is satisfied. We also call NE *p–dominated* if there exists a $\bar{\pi}^p_{-i,a^*} \in \bar{b}^p_{-i,a^*}$ such that the condition in (3) is violated. Finally, note that any element of the joint ε–best response, $\tilde{b}^\varepsilon_{-i}(\pi) = \times_{j \in -i} \tilde{b}^\varepsilon_j(\pi_{-j})$, is a particular element of $\bar{b}^p_{-i,a^*}$ with $p = (1-\varepsilon)^{N-1}$.

### 2.3 Generalised Weakened Fictitious Play

GWFP processes have been analysed using results from stochastic approximations and differential inclusions. Similarly, we make use of these tools to analyse GWFP processes and FFP in the presence of observation noise. To begin, we describe the classical fictitious play process. Let agent *i*'s *historical frequency* of playing $a_i$, be defined as:

$$\sigma^t_{i,a_i} = \frac{1}{t} \sum_{\tau=0}^{t-1} I\{a^\tau_i = a_i\},$$

where $I\{a'_i = a^\tau_i\}$ is an indicator function equal to one if $a'_i$ is the action played by *i* at time τ, and zero otherwise. We write $\sigma^t = \{\sigma^t_{i,a_i}\}_{i \in N, a^i \in A^i}$ for the vector of these beliefs, and $\sigma^t_{-i}$ for the beliefs about all players other than *i*. In classical fictitious play, the chosen action is a best-response to the historical frequencies of all the other players; $a^t_i \in b_i(\sigma^t_{-i})$. Writing $b(\sigma) = \times_{i \in N} b_i(\sigma_{-i})$ for the set of joint best responses, the fictitious play recursion can be restated as the recursive inclusion:

$$\sigma^{t+1} \in \left(1 - \frac{1}{t+1}\right)\sigma^t + \frac{1}{t+1} b(\sigma^t).$$

In [3], the convergence properties of this recursion are analysed using the theory of stochastic approximation of differential inclusions. Building on this, the class of generalised weakened fictitious play (GWFP) processes are characterised in [9]. These are processes that admit a more general belief–updating process and allow δ–best responses to be played by the agents (i.e. weakened best responses). We write $b^\delta(\sigma) = \times_{i \in N} b^\delta_i(\sigma_{-i})$ for the set of joint δ-best responses. In a GWFP process, beliefs follow the inclusion:

$$\sigma^{t+1} \in (1-\alpha^{t+1})\sigma^t + \alpha^{t+1}(b^{\delta^t}(\sigma^t) + M^{t+1}) \quad (4)$$

with $\alpha^t \to 0$ as $t \to \infty$, $\sum_{t \geq 1} \alpha^t = \infty$, and where $\{M^t\}_{t \geq 1}$ is a sequence of martingale differences. For $\alpha^t$ conditions to hold, the sequence must approach 0 slower than $1/t$, i.e. where $\alpha^t = (C_\alpha + t)^{-\rho_\alpha}$ for constants $C_\alpha$ and $\rho_\alpha \in (0,1]$.[3] It has been shown that as $t \to \infty$, trajectories of (4) approximate the differential inclusion:

$$\frac{d}{dt}\sigma^t \in b^\delta(\sigma^t) - \sigma^t. \quad (5)$$

Furthermore, under the assumption that $\delta \to 0$, the stationary points of this differential inclusion are limit points of the difference process, which correspond to a subset of NE to which any GWFP process converges (as shown in [9]). Hence the limit set of a GWFP process (4) is a connected internally chain-recurrent set of the differential inclusion (5), which in turn implies that the limit set of a GWFP process consists of a connected set of NE in potential games, two-player zero-sum games, and generic $2 \times n$ games.

## 3 Games with Perturbed Observations

In this section we introduce a model of perturbed observations in repeated games. Now, standard results on learning in games assume that if players observe their opponents' actions, they do so without noise. A more realistic scenario is that action observations include some noise: this could happen because, for example, noisy sensors are used to detect the other agents' actions, or agents report their actions using a noisy communication medium. After defining the setting, we investigate how relaxing the assumption of perfect observations affects fictitious play processes.

In more detail, although we could use any model of observation noise (e.g. Gaussian), we choose to use the following model of noisy action observations:

**Definition 1** *A game with perturbed action observations is a game in which, when the joint action $a \in A$ is played, each agent $i \in N$ receives the observation of each $a_j$ in $a_{-i}$ with probability $1 - \varepsilon$, and a different, randomly generated action $\tilde{a}_j \neq a_j$ with probability ε.*

Beyond this, we also assume that agents have some knowledge of the process generating observation perturbations,

---

[3] It is useful to think of $\{M^t\}_{t \geq 1}$ as a sequence of zero–mean random perturbations, caused by responses to non–best response profiles, whose effect eventually disappears.

which we feel is reasonable in our application domains. For example, in settings where observation perturbations are generated by noisy sensors, agents typically possess a probabilistic model of their sensors. Similarly, agents employing noisy communication channels to coordinate usually have some knowledge of the way that errors are generated in their messages. Furthermore, these noise–generating models do not need to be precise, because, as we show later, our results apply even with adaptive noise modelling functions as long as the errors in those functions are bounded.

We now show that perturbations to action observations prevent standard GWFP processes from converging, using the concept of $p$–dominance and results from stochastic approximations. In general, the non–convergence of all GWFP processes in games with perturbed action observations can be explained by examining their effects on the belief updates given in (4).

**Theorem 1** *Perturbations to action observations with noise level $\varepsilon$ will prevent a GWFP process from converging to a $p$–dominated NE for $p < (1-\varepsilon)^{N-1}$ in any class of games in which standard GWFP processes do converge (as listed in Section 2.3).*

**Proof:** Recall that for a GWFP process to converge, $b^\delta(\sigma) \to b(\sigma)$ as $t \to \infty$. If this assumption is not satisfied, the difference process defined by GWFP does not have any limit points, and, furthermore, in general, its trajectory will not be contained within a particular sub–region of the joint strategy space. Given this, in the presence of noise in action observations, beliefs follow an inclusion given by:

$$\sigma^{t+1} \in (1-\alpha^{t+1})\sigma^t + \alpha^{t+1}(\tilde{b}^\varepsilon(\sigma^t) + M^{t+1}), \quad (6)$$

where $\tilde{b}^\varepsilon(\sigma^t) = \times_{i \in N} \tilde{b}_i^\varepsilon(a_{-i})$, which captures the effects of action observation noise. Note that, unlike the noise–free situations in which standard GWFP processes are assumed to operate, perturbations to action observations do not disappear. The persistence of this noise in the action observations implies that the trajectories of (6) are stochastic approximations of a new differential inclusion, given by:

$$\frac{d}{dt}\sigma^t \in \tilde{b}^\varepsilon(\sigma^t) - \sigma^t. \quad (7)$$

Stationary points of (7) exist only at $(1-\varepsilon)^{N-1}$–dominant equilibria, rather than at all NE. Furthermore, if a stationary point exists, the sequence of martingale differences, $\{M^t\}_{t \geq 1}$, also converge to zero as play converges to this stationary point, because the frequency of non–best response actions declines as play converges. Therefore, in the presence of perturbed observations of actions, standard GWFP processes converge only to $p$–dominant equilibria for $p = (1-\varepsilon)^{N-1}$. □

**Corollary 1** *Standard GWFP processes are not generally guaranteed to converge in games with perturbed action observations.*

**Proof:** This is a consequence of Theorem 1: If all of the NE in the game are $p$–dominated for $p = (1-\varepsilon)^{N-1}$, then there must exist an action $a_i \neq a_i^*$ for some player such that:

$$r_i(a_i^*, \sigma_{-i}^t) - r_i(a_i, \sigma_{-i}^t) < 0.$$

because $\tilde{b}_{-i,a^*}^\varepsilon \geq \sigma_{-i,a^*}^t$. As a consequence, for high enough levels of noise in action observations, the differential inclusion in (7) may not have any stationary points, the sequence $\{M^t\}_{t \geq 1}$ does not disappear, and standard GWFP processes are not guaranteed to converge. □

Here we note that, although conceivably arising in games with perturbed action observations, no other forms of attractors, such as correlated equilibria or chaotic attractors, are either (i) analytically proven to be convergence points for GWFP, or (ii) appropriate for the optimisation problem we intend to apply the technique to, in which optimal configurations of agents' actions arise at NE of the game.

An example of a game in which a standard GWFP does not converge in the presence of perturbed action observations is given below for our UAV target tracking domain.

**Example 1** *Consider two UAVs employing GWFP to coordinate on positions that maximise sensor coverage of a target. The optimal configuration has one UAV above the target and the other in a secondary point, giving rewards of $(1,0)$ or $(0,1)$. If both find secondary points then they get no reward $(0,0)$, while if both attempt to get above the target they collide, giving $(-4,-4)$. This is shown below:*

|          |           | Player 2  |       |
|----------|-----------|-----------|-------|
|          |           | Secondary | Above |
| Player A | Secondary | 0,0       | 1,0   |
|          | Above     | 0,1       | -4,-4 |

*This is a potential game, so in the absence of noise in action observations, UAVs coordinating via GWFP will converge to one of the NE, which are located at either of the off–diagonal pure actions, and a symmetric mix with $0.8$ probability of choosing the secondary position by both UAVs.*

*However, if the UAVs' action observations are perturbed with probability $\varepsilon > 0.2$ (because, e.g., they cannot accurately detect each others' movements), then a GWFP process will not converge. To see this, note that the expected rewards to the row UAV for either action are equal at the mixed strategy equilibrium. As such, even if the column UAV continually plays the* secondary *position, if $\varepsilon > 0.2$ then the row UAV will not take the* above *position, because it observes the other UAV choosing the* above *position with too great a probability. Noise in action observations directly prevents the UAVs' from converging to the mixed NE.*

In practice, GWFP processes do converge to a strict NE in the presence of noise until $\varepsilon > 1 - p^{\frac{1}{N-1}}$ for the minimum $p$–dominant equilibrium in the game. Beyond this noise level, the process never stays at a NE. We demonstrate this type of threshold behaviour in Section 4.3.

# 4 Filtered Fictitious Play

In this section we introduce filtered fictitious play (FFP). The section is divided into three parts. In the first, we derive a general method for integrating a noise filter into the GWFP belief revision process. This method allows us to analyse the trajectories of any FFP process using the same techniques as GWFP. As a concrete example of an FFP process, we derive a specific version of FFP that makes use of a Bayes filter on the noise in action observations, which is appropriate to settings where the noise generating process is known to the agents. In the second part of the section, we assume that the agents have access to a noise filtering process of bounded precision $\eta$, and present our main result: FFP will converge to any $p$–dominant NE for $p$ up to $(1-\eta)^{N-1}$. A corollary is that when the noise generating process is known to the agents, a Bayes filter gives accurate estimates of the true frequency of actions played, therefore FFP converges to any strict NE and will converge in any game that GWFP processes converge to a strict NE (i.e. the class of potential games). The final part of this section illustrates the improved convergence properties of FFP with a Bayes noise filter by comparing it to a naïve, standard GWFP process in the game from Example 1.

## 4.1 Filtered Belief Revision

FFP uses a filter to remove noise in action observations, while operating within the general analytical framework provided by GWFP. As such, we can prove its convergence in many games with perturbed action observations. To understand why this is necessary, see that, from (6), in standard GWFP processes, a noisy observation alters an agent's belief in player $i$'s action $a_i$ in the following way:

$$\sigma_{i,a_i}^{t+1} = \sigma_{i,a_i}^t + \alpha^{t+1}(P(\tilde{a}_i^t|a_i^t) - \sigma_{i,a_i}^t)$$

where $P(\tilde{a}_i^t|a_i^t) = \tilde{b}_{i,a_i}^\varepsilon(\sigma_{-i})$. Specifically, GWFP naïvely updates its belief in the observed action with the probability that what is observed is the true action, $P(\tilde{a}|a_i)$, and the same *mutatis mutandis* for unobserved actions.

This is incorrect. We need the probability that $a_i$ is played, given the observation $\tilde{a}_i$. In general, we can use a noise correcting filter to generate this probability,[4] i.e.:

$$P(a_i|\tilde{a}_i) = \phi(\tilde{a}_i, P(a_i), \theta),$$

where $\phi$ takes the observation, a set of filter coefficients $\theta$ and (possibly) prior information $P(a_i)$ as parameters. By setting the prior to the current belief, $P(a_i) = \sigma_{i,a_i}$, we then intergrate the filter into the best response correspondence, which results in the following belief update rule:

$$\sigma_{i,a_i}^{t+1} = \sigma_{i,a_i}^t + \alpha^{t+1}\left(b_i(\phi(\tilde{a}_i^t, \sigma_{-i}^t, \theta)) - \sigma_{i,a_i}^t\right) \quad (8)$$

[4] We leave the investigation of the practicalities of this interesting avenue of research for future work, and focus here on the general derivation.

We assume that the filter in use is able to remove noise and return an estimate of precision $\eta$. To define this formally, the 'true' (i.e. unperturbed) belief in $i's$ actions is given by:

$$\bar{\sigma}_i^t = \frac{1}{t}\sum_{\tau=0}^{t-1} I\{a_i^\tau = a_i\},$$

We say that the maximum precision of any belief revision process is $\eta$ if, given a fixed action profile $a_{-i}$:

$$\lim_{t\to\infty} |\bar{\sigma}_i^t(a_{-i}) - P(a_{-i}^t|\tilde{a}_{-i}, \sigma_{-i}^{t-1}, \theta)| = \eta \quad (9)$$

In the next section, we make use of this bound to characterise the class of games in which FFP converges to a NE.

However, before moving on to our main results, we now build on this general method to derive a belief revision rule for situations where the noise generating process is known (e.g. from a probabilistic sensor model or known properties of the communication medium). In such circumstances, we can use Bayes rule to compute $P(a_i|\tilde{a}_i)$, that is:

$$P(a_i|\tilde{a}_i) = \frac{P(\tilde{a}_i|a_i)P(a_i)}{\sum_{a_i'\in A_i} P(\tilde{a}_i|a_i')P(a_i')}$$

Again, we set the prior to the current belief, $P(a_i) = \sigma_{i,a_i}$, but now make use of the the noise generating function, $\tilde{b}^\varepsilon(a_{-i})$, as defined in (1), to derive the following filtered best response correspondence:

$$b(P(a_{-i}^t|\tilde{a}_{-i}^t, \sigma_{-i}^t, \varepsilon)) = \frac{\tilde{b}_{i,a_i}^\varepsilon(\sigma_{-i}^t)\,\sigma_{i,a_i}^t}{\sum_{a_i'\in A_i}\tilde{b}_{i,a_i'}^\varepsilon(\sigma_{-i}^t)\,\sigma_{i,a_i'}^t}. \quad (10)$$

The result is a *Bayesian* filtered belief revision rule:

$$\sigma_{i,a_i}^{t+1} = \sigma_{i,a_i}^t + \alpha^{t+1}\left(b(P(a_{-i}^t|\tilde{a}_{-i}^t, \sigma_{-i}^t, \varepsilon)) - \sigma_{i,a_i}^t\right) \quad (11)$$

Finally, note that the above belief revision specifications are sufficient if the goal is to find a pure NE, but at this stage it is unclear if it is sufficient for convergence to mixed NE, although this is not likely to be the case.[5]

## 4.2 Convergence to Nash Equilibrium of FFP

We now prove that the FFP process defined in (8) converges to strict NE in many games in which perturbations to action observations prevent the convergence of standard GWFP processes. The result depends on proving that as $t \to \infty$, the filtered belief update process comes closer to the 'true' belief corresponding to a $p$–dominated NE than a standard GWFP process, and uses the definition of precision in (9).

[5] C.f., several authors have investigated using Dirichlet processes to update distributions over mixed strategies in an attempt to capture convergence to mixed NE in games with perfectly observable actions, but without success (see, e.g. [6, 15]).

**Theorem 2** *FFP with belief revision of maximum precision $\eta$ can converge to a $(1-\eta)^{N-1}$–dominant equilibrium.*

**Proof:** This theorem follows from the same reasoning as Theorem 1, where now the lower limit on $p$ is related to the precision of the belief revision process. As such, the stationary points of the stochastic approximation:

$$\frac{d}{dt}\sigma^t \in b(P(a^t_{-i}|\tilde{a}_{-i}\sigma^t_{-i},\varepsilon)) - \sigma^t \quad (12)$$

only exist at $(1-\eta)^{N-1}$–dominant equilibria, where $\eta$ is the precision of the estimator of $P(a^t_{-i}|\tilde{a}_{-i}\sigma^t_{-i},\varepsilon)$. □

Thus, if the process generating perturbations is known, then using the Bayesian filtered belief revision rule in (10), ensures that $\eta < \varepsilon$, giving us the following corollary.

**Corollary 2** *If the perturbation process is known, then FFP using the Bayesian filtered belief revision will converge to a strict NE in all games with action observation perturbations in which standard GWFP converges to a strict NE in the absence of noise.*

Most importantly for the application of FFP to distributed optimisation problems, this includes potential games. In practice, however, very high levels of noise will prevent any FFP from converging.

On the speed of convergence of beliefs in FFP processes: Although using filtered updates and $\lambda^t = 1/t$ may cause beliefs to move more slowly than classical fictitious play in games with low noise, this can be significantly boosted by putting more weight on the newest filtered updates, by altering the form of $\lambda^t$. Recall that one convergent form for this sequence is $\lambda^t = (C_\lambda + t)^{\rho_\lambda}$, with $C_\lambda \geq 0$ and $\rho_\lambda \in (0,1]$. By setting $\rho_\lambda$ closer to zero, more weight is given to the filtered update of the belief. This improves the speed which $\phi(\cdot)$ filters noise. However, for the process to still converge, the players' beliefs must not be completely updated with the filtered revision (i.e. $\rho_\lambda$ cannot be set to 0).

### 4.3 Numerical Demonstration

We now demonstrate the non-convergence of GWFP processes in the simple UAV monitoring game with perturbed action observations, as described in Example 1, and also show how FFP processes do converge in this same domain. We do this to demonstrate both the brittleness of GWFP in the presence of noise and robustness of FFP with that same noise. This evaluation give a concrete demonstration of the previous two theoretical results.

Results are shown in Fig 1, with error bars showing two standard errors. GWFP initially converges to the NE nearly all of the time (%$N > 96$), however, for $\varepsilon \geq 0.2$, its performance quickly drops to 0. This is because it cannot learn an accurate mixed strategy for the other agent since it sees very noisy observations and assumes that this represents the mixed strategy being played. In contrast, FFP is much more resistant to noise and is unaffected until $\varepsilon \geq 0.4$. At this point, FFP also drops to 0 because there is very little information in the observations, so effectively, the agents cannot observe what the other is playing at all. In summary, FFP has a convergence rate of 82% whereas GWFP has only 41% across the range of $\varepsilon$. These results show the power of using Bayesian inference to account for observation noise — and consequently show that FP may still be used in highly uncertain domains. Next, we show how FFP is used in a learning algorithm for POSGs in which agents are uncertain about the current stage game.

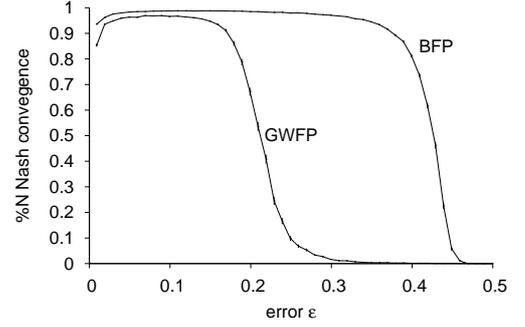

**Figure 1:** Convergence in the UAV monitoring game.

## 5 Sequential Decision–Making

Now that we have proved FFP's convergence in single–state games with perturbed action observations, we can use FFP as a component of an algorithm for solving Dec–POMDPs (i.e. team POSGs). We concentrate on the specific sub-class where agents know the utility and transition functions in advance. The problem here is to learn a best action in each state and furthermore, accurately track which stage game is being played. In the first part of this section we derive an algorithm using FFP for this problem, and then in the second part we demonstrate its utility in a standard Dec–POMDP, namely, cooperative box pushing.

### 5.1 FFP–based Lookahead Search Algorithm: LFFP

Our algorithm, LFFP, is designed for POSGs in which agents jointly optimise a potential function, and in which the agents' actions are the only cause of transitions between stage games. In the case of a team game, where the agents directly optimise a potential function given by the global target function, this class of POSGs corresponds to a sub–class of Dec–POMDPs with action dependent state transitions. In these games, tracking the current state is very important because the current stage game influences the rewards for different actions, and, furthermore, agents must be able to predict state transitions in order to plan into the future to achieve the optimal reward.

Formally, a POSG comprises a finite set of stage games, $S = \{\Gamma^1, \ldots, \Gamma^{|S|}\}$, where each stage game is a noncoop-erative game as described in Section 2.1, with actions $A_i^s$ and utilities $u_i^s(a^s)$ $\forall i$. Movement between stage games is controlled by a state transition function $T : S \times A \to S$. The total rewards a player earns is the sum of rewards from all future stage games, discounted by $\gamma$. States are partially observable because noise in the player's observations of $a_{-i}$ induces a distribution over $T$.

LFFP is an online lookahead search algorithm (similar to [19]) that uses the likelihood's of other agents' actions to track the current game being played, and searches into the future for a limited horizon. In order to return consistent actions we assume that an agent knows the set of actions it individually can take in a given state, but must infer the actions for the other agents from its beliefs about the current state. LFFP uses FFP's belief revision to update beliefs and, therefore, also to track the state — we later show that using standard GWFP to do this results in very poor performance because it does not accurately track the state.

Specifically, each agent computes an action to play in state $s$, given its beliefs over others' actions $\sigma^s$. This is done by evaluating a recursive expression, which uses the knowledge of the transition function and the observations received so far, along with the agent's current possible action set to construct a probability of a particular stage game being played in the furture. Now, in order to explore and learn the payoff structure of the problem, the agent randomises between an *optimal* evaluation given its belief and an *optimistic* evaluation, where it assumes that the other agents act to maximise its reward in the next time step.[6] The optimal action evaluation in state $s'$ at $d$ steps into the future, $V^*(d, s')$, is used with probability $1 - \xi$, and is given by:

$$V^*(d, s') = \max_{a_i \in A_i} \sum_{a_{-i} \in A_{-i}} \Big( \prod_{j \in -i} \sigma_j^{s'}(a_j) \Big) \Big( r_i^{s'}(a_i, a_{-i}) + \gamma \sum_{s'' \in S} T(s'', a_i, a_{-i}, s') V(d-1, s'') \Big)$$

where $V(d-1, *)$ are the values, optimal or optimistic, computed at each of the nodes at level $d-1$ above $(d, s')$.

The optimistic value of an action, $V^{opt}(d, s)$, is used with probability $\xi$, and is given by:

$$V^{opt}(d, s) = \max_{a_i \in A_i} \max_{a_{-i} \in A_{-i}} \Big( r_i^s(a_i, a_{-i}) + \gamma \sum_{s' \in S} T(s', a_i, a_{-i}, s) V(d-1, s') \Big)$$

These values are propagated down the search tree, beginning at the leaves $D$ steps into the future, with the agent choosing to use the optimal or optimistic evaluation at each decision node. Finally, at the root node, the $a_i \in A_i$ that maximises $V^*(0, s)$ is chosen (i.e no randomisation is used at the root node). The $\xi$ is annealed over time, following $\propto 1/t^{(1/2)}$, where $t$ is the iteration of the algorithm, so that in the limit only $V^*(d, s')$ are evaluated.

### 5.2 Numerical Evaluation of LFFP

In this section, we evaluate the performance of LFFP for POSGs in the canonical cooperative box pushing problem [17]. This problem consists of a gridworld containing two agents, two small boxes and a large box. The agents can independently push the small boxes into the goal to get a small reward, but the highest reward is if both agents together push the large box into the goal, which neither can do on its own. The agents are penalised for taking too long or bumping into each other or the walls of the world. Agents can move forward a step, turn left or right or stay still. Finally, they can only see the state of the other agent if it is nearby. The game consists of 100 states, 4 actions and 5 observations, making it a challenging problem.

Specifically, agents are not supplied with the stage game that they are playing, and further to this, the only positive reward in the game is for pushing the boxes into the goal area. Consequently, any solution needs to accurately estimate which stage game is being played and also plan a path to the goal state. There are few algorithms which are capable of finding a NE in such problems with uncertainty that are also capable of scaling to more than two agents — for example, most centralised offline Dec–POMDP models suffer from severe complexity bottlenecks. In contrast, LFFP does scale, since it effectively regards the other agents as a single entity and learns a best response to their joint action, and furthermore, avoids estimating the joint history such as PBPG [21].

In this experiment, we compare LFFP to online local search using the standard GWFP update rule (which we call LGWFP), and the current state of the art in approximate offline Dec–POMDP planners, PBPG. LFFP and LGWFP are online learners whilst the benchmark is an offline planner, so we compare the reward accumulated for our learned policy after 100000 timesteps to the planned policy from PBPG over the same horizon (100 timesteps). PBGP allows the specification of the number of points so we use 100 as per the published algorithm. As a further note, the run–time for the offline planning is comparable to the learning time we allow LFFP and LGWFP.

As Fig 2 shows, LGWFP performs more poorly in this game than GWFP did in the UAV monitoring game earlier. Specifically, it stops converging even sooner than in Fig 1 ($\epsilon \geq 0.15$). This is because the error in estimating other agents' mixed strategies is compounded through the search resulting in both poor stage game estimation and conver-gence. In contrast, LFFP does nearly as well as the normal game formalism. In more detail, it only stops converging for $\varepsilon \geq 0.35$, which is a very high level of noise. The slight reduction in performance is because again the error in random observations starts to be compounded by chaining multiple observations together. However, that said, these results show that LFFP is a useful multi-agent learning strategy in stochastic games. The benchmark PBGP only performs as well as our algorithm for a very small amount of noise, and after this it even fails to find the non-cooperative solution (each agent pushes a small box). One possible reason for this is because more noise requires more points to sample, whilst we kept this parameter fixed at the value used by the authors themselves — and increasing this value causes the runtime to increase dramatically. This is to be contrasted with LFFP, which when it cannot find a good model of the other agents, defaults to a safe individual policy, which may be lost in point based approaches.

---
[6]The NashQ algorithm [10] uses a similar bias towards *coordinated equilibria*, which are those NE that maximise all agents' rewards. Now, because Dec-POMDPs are team games, there exists a coordinated equilibrium, and we bias exploration towards it at each evaluation node in the search tree by making the described optimistic evaluation.

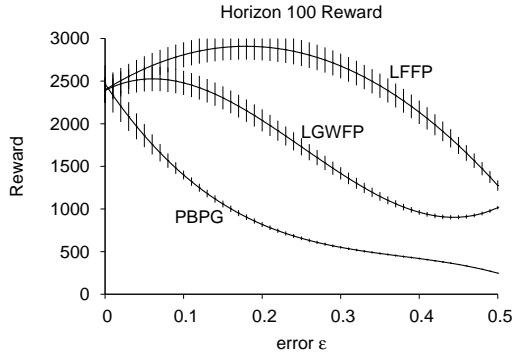

**Figure 2:** Rewards in the cooperative box pushing problem.

## 6 Conclusions

In this paper, we demonstrated that iterative learning, which is computationally cheap, can be applied to static optimisation and sequential decision–making with perturbations to action observations. Specifically, we introduced filtered fictitious play, which can learn NE in potential games when action observations are perturbed by noise. As such, it is suitable for distributed optimisation problems with noisy observations of actions or communication channels. We proved that standard fictitious play processes do not, in general, converge under perturbed action observations, and demonstrated FFPs benefits over GWFP in a noisy coordination game. We then used FFP to derive a solver, LFFP, for Dec–POMDPs in which state transitions are a function of agents' actions, and uncertainty arises as a result of noise in action observations. We showed the utility of LFFP in a partially observable stochastic game — the cooperative box pushing problem. Here LFFP showed significantly resistant to noise in observations and consistently outperformed the state of the art in this setting.

Now that we have proven the convergence of FFP and demonstrated the efficacy of LFFP for POSGs and Dec–POMDPs, future work will thoroughly analyse the convergence and complexity of LFFP. With this established we will investigate if the assumptions under which LFFP operates can be relaxed to allow it to learn the structure of the underlying POSG (i.e. rewards and transitions) as well as the policies of the other agents.